\documentstyle[12pt]{article}
\begin{document}
\begin{center}
\begin{large}

{\bf On the Belitz-Kirkpatrick comment on "Specific heat of a Fermi system
near ferromagnetic quantum phase transition", by I.Grosu, D.Bodea and
M.Crisan (cond-mat/0101392)}
\end{large}
\vspace{1cm}

{\bf M.Crisan, I.Grosu and D.Bodea}
\vspace{0.5cm}

{\bf Department of Theoretical Physics, University of Cluj, 3400 Cluj, Romania}
\end{center}
\vspace{1cm}

In a recent comment [1] Belitz and Kirkpatrick made some pertinent observations
concerning our calculations [2] of the specific heat in a quantum phase
transition which appear in a itinerant-electron ferromagnet. We mention from
the beginning that these observations are not to defend our paper but to 
make more clear the problem from our paper which is not the same with
the new points of the authors. However, we mention
that indeed $C_{V}\sim T\ln T$ has been obtained first by Millis [3] for the
marginal case $z=d$ (recently it was showed [4] that it could also appear near
a Lifshits point), but in this case we belive that it is given by the presence
in the theory of the dangerous irrelevant parameter which leads to the 
logarithmic correction as was mentioned recently by Belitz et al.[5] on the 
basis of scaling analysis. Before considering the specific feature of our
model and calculations we will start with a short analysis of the experimental
results, and the theoretical models proposed in literature for the
quantum phase transition (QPT) in the itinerant-electron ferromagnetism.
\vspace{0.5cm}

{\bf A. A short review of experimental data and the phenomenological
approach}
\vspace{0.5cm}

It is well known that that quantum phase transitions in clean itinerant-electron
ferromagnet (we will use clean and disordered in the same sense as Kirkpatrick
and Belitz [5] and Vojta [6]) appear in a small number of systems 
(if we refer to the antiferromagnetic QPT) as
$MnSi$ or $Th_{1-x}U_{x}Cu_{2}Si_{2}$ and this transition is given by doping
or preassure. One of the most remarcable property of these materials is the
non-Fermi behavior of the transport and magnetic properties. Pfleiderer et al.
[7] formulated a phenomenological Moryia-like model for the explanation of the
non-Fermi behavior of $MnSi$, the main point being the state $d=z=3$ which
is a non-Fermi liquid. This is in agreement with the Millis [3] analysis from
Renormalization Group (RG) calculation, and called marginal case.
\vspace{0.5cm}

{\bf B. Renormalization Group approaches}
\vspace{0.5cm}

The first way to use the Renormalization Group (RG) method
, due to Hertz [10] was reconsidered by Millis [3] and recently
by Lavagna [11] and used for the clean itinerant-electron paramagnetic phase.
The susceptibility $\chi(\vec{q}, \omega)$ given by Lindhard function arround
$q=\omega=0$ is written as:
$$\chi^{-1}(\vec{q},\omega)=\chi^{-1}(0,0)-b\left(\frac{q}{k_{F}}\right)^{2}+
ia\frac{\omega}{q v_{F}}\eqno{(1)}$$
The imaginary term can be written as $\omega/\Gamma(q)$, where
$\lim_{q\rightarrow 0}\Gamma(q)=0$, 
because the fluctuations of the order parameter
in ferromagnetic order satisfies the conservation law. The RG theory is 
developed using a $\phi^{4}$-model for the fluctuations of the order
parameter, and for $d_{eff}=d+z>4$ the interaction becomes an irrelevant
parameter which leads to a Gaussian model. Studing the crossover from quantum
to classical behavior it was showed (see Ref.[3,4]) that in the specific heat the
$T\ln T$ is given by the quantum contribution. 

The second point of view was
recently pointed out by Belitz, Kirkpatrick and Vojta (BKV) in a relevant
number of papers [5,6] (the starting papers are cited in these references) who
showed the non-analyticity of $\chi(\vec{q}, \omega)$ and for $d=3$, and 
at finite temperature we have:
$$\chi^{-1}(\vec{q},\omega)=\delta+c_{3}\left(\frac{q}{2k_{F}}\right)^{2}
\ln\frac{2k_{F}}{q}+c_{2}q^{2}+\frac{|\omega|}{\Gamma(\vec{q})}\eqno{(2)}$$
where $\Gamma(\vec{q})\sim q$. This form gives a first order phase transition
and leads to the non-analyticity of the Landau-Ginsburg theory that is given by
the singular behavior of the coefficients of this expansion. This singular
behavior is due to the integrating out of the soft 
mode which is in this case
particle-hole excitations mode.
In fact this appear when $\chi(\vec{q},\omega)$ is calculate using a linear
dispersion for the electron-hole excitations. The Lindhard function 
$\chi(\vec{q},\omega)$ is calculated using the Green functions with quadratic
wave vector dependence. Recently, the authors [12] developed a local field
theory (a theory with a definite Landau-Ginsburg action when $q\rightarrow 0$,
$\omega\rightarrow 0$) and for a disordered ferromagnet, keeping the soft modes
which in this case are diffusive. The logarithmic correction to the scaling was
presented as due to the Wegner corrections. The problem of the clean 
itinerant-electron ferromagnetic transition was not treated using the local
field theory, but it seems to us that a similar RG treatment can be done.
\vspace{0.5cm}

{\bf C. Model proposed in our paper [2]}
\vspace{0.5cm}

Our paper started from a very special class of systems $F_{x}Pd_{1-x}$ (where
$F$ is a ferromagnetic impurity, and $x$ is the concentration) first studied
by Doniach and Wohlfarth [8] and recently by Nicklas et al.[9] which showed
that such a system becomes non-Fermi. These systems are very different from the
systems mentioned above because $Pd$ is a paramagnetic metal which is close to
the Stoner instability and a small concentration of magnetic impurities drive
this metal in a ferromagnetic state, but near a critical concentration
associated with the quantum phase transition this system becomes also non-Fermi.
This behavior is difficult to be explained in terms of disorder, and we adopted
the idea from Ref.8, that a strong polarization of the Fermi liquid 
couples to the fluctuations of the magnetic impurities. The susceptibility
of this system has the form:
$$\chi^{-1}(\vec{q},\omega)=\delta-aq^{2}-b\frac{\Delta\omega}{q^{2}}-
i\frac{\omega}{\Gamma\cdot q}\eqno{(3)}$$
where $\delta$ is the concentration dependent distance from the quantum 
critical point. Starting with these considerations we will discuss the possible
theoretical models which can give for the specific heat a $T\ln T$ behavior,
which demonstrate the non-Fermi character of the system. First we have to define
the fluctuations of magnetic impurities. Certainly this is not a critical mode
and have a strong local character. This gives a very special form for
$\chi(\vec{q},\omega)$ approximated by Eq.(3), which for $\Delta=0$ gives the
behavior similar with Eq.(1). This equation is different from Eq.(2) because
of the absence of a $(q/2k_{F})^{2}\ln(2k_{F}/q)$ term in the BKV model. 
On the other hand we mention that due to the electron-spin coupling we
cannot integrate out the soft mode (electron-hole mode) and this is the
explanation why Eq.(3) is similar to Eq.(1), if $\Delta=0$.
>From the RG scaling equation we have to get $\Delta\neq 0$ a relevant parameter,
and $\Gamma=const$. This can be obtained if $z<4$ and $z=3$. This is specific
for the studied system and gives to $u$ (interaction) the character of 
dangerous irrelevant parameter. 
A sistem with
$z=2$ still keep valid the condition $d_{eff}=d+z>0$ which can give a $T\ln T$
in the specific heat. It is easy to see that this behavior is associated with
the specific model defined by Eq.(3), which will never contain a non-analytic
contribution because we used the Lindhard form for $\chi(\vec{q}, \omega)$
of the electrons. Finally, we mention that even for a more eleborated model, the 
non-Fermi behavior, mentioned in fact by Belitz et al.[14], have to be reobtained
for the QPT in the itinerant-electron systems in $d=3$.
\vspace{1cm}

{\bf References}

[1] D.Belitz, T.R.Kirkpatrick, cond-mat/0102064

[2] I.Grosu, D.Bodea, M.Crisan, cond-mat/0101392

[3] A.J.Millis, Phys.Rev.B 48, 7183, (1993)

[4] C.P.Moca, I.Tifrea, M.Crisan, Phys.Rev.B 61, 3247, (2000)

[5] T.R.Kirkpatrick, D.Belitz, cond-mat/9707001 (see also the references

therein concerning this problem obtained by Belitz, Kirkpatrick, Vojta

(BKV))

[6] T.Vojta, Ann.Phys.(2000) (to be published)

[7] C.Pfleiderer, G.J.McMullan, S.R.Julian, G.G.Lonzarich, Phys.Rev.B 

55, 8330, (1997)

[8] S.Doniach, E.P.Wohlfarth, Proc.Royal.Soc.(London) 296, 442, (1967)

[9] M.Nicklas, M.Bredo, G.Kenbel, F.Mayr, W.Trinke, A.Ladl, Phys.Rev.Lett.

82, 4286, (1999)

[10] J.A.Hertz, Phys.Rev.B 14, 1165, (1976)

[11] M.Lavagna, cond-mat/0102119

[12] D.Belitz, T.R.Kirkpatrick, M.T.Mercaldo, S.L.Session, cond-mat/0008061

[13] D.Belitz, T.R.Kirkpatrick, M.T.Mercaldo, S.L.Session, cond-mat/0010377

[14] D.Belitz, T.R.Kirkpatrick, R.Narayanan, T.Vojta, Phys.Rev.Lett.85,

4602, (2000)

\end{document}